\RequirePackage[2020-02-02]{latexrelease}
\documentclass[superscriptaddress,secnumarabic,
amssymb,amsmath,nobibnotes,aps,prd,showkeys,showpacs,nofootinbib]{revtex4}%
\usepackage{subfigure}
\usepackage{epsf}
\usepackage{epsfig}
\usepackage{graphicx}
\usepackage{tabularx}
\usepackage{hhline}
\usepackage{subfigure}
\usepackage{subcaption}
\usepackage{wrapfig}
\usepackage{lipsum}
\usepackage{epstopdf}
\usepackage{epsf}
\usepackage{epsfig}
\usepackage{scrtime}
\usepackage{multirow}
\usepackage[multiple]{footmisc}
\usepackage{amsfonts, graphics}
\usepackage{bm}
\usepackage[utf8x]{inputenc}
\usepackage{amsmath}
\usepackage{amsfonts}
\usepackage{amssymb}
\usepackage{color}
\usepackage{natbib}%
\providecommand{\U}[1]{\protect\rule{.1in}{.1in}}
\newcommand{\be}{\begin{equation}}
\newcommand{\ee}{\end{equation}}

\newcommand{\mincir}{\raise
-3.truept\hbox{\rlap{\hbox{$\sim$}}\raise4.truept\hbox{$<$}\ }}
\newcommand{\magcir}{\raise
-3.truept\hbox{\rlap{\hbox{$\sim$}}\raise4.truept\hbox{$>$}\ }}

\begin{document}

\title{Herrera Complexity and Shadows of Spherically Symmetric Compact Objects}

\author{Subhasis Nalui}
\email{subhasis.rs@presiuniv.ac.in}

\author{Subhra Bhattacharya}
\email{subhra.maths@presiuniv.ac.in}
\affiliation{Department of Mathematics, Presidency University, Kolkata-700073, India}


\begin{abstract}
 In this work we investigate the effect of complexity factor on the formation of photon spheres for spherically symmetric compact objects. The complexity factor obtained from the orthogonal splitting of the Riemann curvature tensor connects the geometric attributes of a compact spherically symmetric gravitating object with its matter inhomogeneity and pressure anisotropy via a scalar term. The novelty of the complexity factor is the inherent simple definition that identifies the evolution of matter tensors inside a given region of space-time. Such identification helps to obtain an equivalence class of gravitating compact objects based on their degree of complexity with zero complexity identified as the simplest system. On the other hand shadows and photon rings have become essential for identifying compact regions of space time characterised by massive gravity. Advanced observational data analysis tools augments the hope for identification of exotic gravitational objects, like the so called ``black hole mimickers" and may serve as testing ground for other gravity theories. In this context we explore how complexity of compact objects (a fundamentally theoretical classification) is connected to the photon ring (an astrophysical observable in the universe) and its stability. We consider zero complexity systems and discuss its significance with respect to (wrt) formation of photon rings and hence shadows. 
 \end{abstract}


\maketitle

\section{Introduction}
\label{sec-introduction}

Symmetric compact structures in the universe are important source of information, that serve as objects for testing general relativity (GR). It is natural that research on these objects are of paramount importance. The key factor in GR research is the link between theoretical formulations and observational evidence. Several mechanism has been devised to theoretically characterise the observable structures in the universe. In this context the complexity factor parametrization \cite{lh1,lh2} of compact objects have recently gained importance. The complexity factor of a self gravitating system is described by the anisotropy in pressure and inhomogeneity in matter density. It is imaginable that complexity of any structure in the universe is defined by the distribution of fluid in a given region. Hence inherent properties of a fluid material like pressure anisotropy and matter density distribution will play crucial role in defining the geometry and gravitational impact of the body, thus justifying their role in defining the ``complexity". Further, such a description of complexity is done through a scalar term obtained from the orthogonal splitting of the Riemann curvature tensor. They have thus become simple methods of classifying compact objects based on their matter distribution. Since its conceptualization they have been used in several contexts \cite{cf}.

The massive gravity regions of space time are important because they are ideal for testing GR. A massless photon travelling along the null geodesic of a massive gravity object can get captured at a certain critical energy and radius, where it revolves around the object to form a sphere of light or the photon sphere. The region inside the photon sphere appears dark in the two-dimensional visual frame of a distant observer giving the impression of shadow or image of the object. The stability of these photon spheres and their differential lensing property are important markers to identify and distinguish between different massive gravity regions. The critical radius for a photon gravitationally confined to revolve around some massive object was given by Synge in 1966 \cite{syn}. Consequently in 1974 and 1979, works by Page and Thorn \cite{pt} and Luminet \cite{lm} respectively, lead to the development of properties of thin accretion disc surrounding a massive black hole. Since then numerous literature have perfected the theory of shadows and photon spheres by black holes \cite{bh7, bh8, bh6, bh1, bh2, bh3, bh4, bh5} and other massive gravity exotic objects like the wormhole  \cite{wh1,wh2, wh3, wh4, wh5,wh6, bh6}, naked singularity \cite{bh7, ns1, ns2,ns3} and more recently gravastars \cite{gs2,gs15,gs16}. Finally almost five decades after the initial calculations on existence of black hole image were made, gravity scale images of of black holes at the centre of the Messier 87 galaxy and our own Milky Way galaxy, the M87* and Sgr A* were drawn using the Event Horizon Telescope (EHT) data \cite{eht1, eht2}, thus opening up a new era of precision observations in general relativity (GR) and fundamental physics. Testing not just massive objects within the realm of GR, it also opened up avenues for testing other gravity theories and exotic objects such as those mentioned before \cite{sv,go,kp,rs1,mg,vd,nm,nt}. 

In this context we aim to mathematically relate the complexity of a region of space time with their corresponding existence of photon spheres or antiphoton spheres.  We specifically postulate our results with respect to the black hole and wormhole like space time. We try to relate the positive complexity, negative complexity  and zero complexity with the existence or absence of the photon sphere. 

The paper is organised as follows: In section 2 we provide a brief review of the complexity factor, in section 3 we describe in brief the relevant equations to obtain and analyse the existence of photon spheres. In section 4 we mathematically relate the complexity of a space time with the corresponding effective potential used to find the photon sphere. In sections 5 and 6 we use the black hole like space time and the Morris Thorne Traversable wormhole space time and relate their complexity with their corresponding existence of photon spheres. In section 7 we analyse the zero complexity scenario while in section 8 we discuss our results in the context of a general space time. We finally show the results using graphical representations of the photon sphere, shadow and complexity factor. Finally we end with a brief discussion in section 9.

\section{Complexity factor: A brief description}
In this section, we will discuss about the complexity of static spherically symmetric geometric structure expressed by the line element:
\begin{equation}
    ds^{2}=-A^{2}dt^{2}+B^{2}dr^{2}+C^{2}(d\theta^{2}+sin^{2}\theta d\Phi^{2}),\label{metric1}
\end{equation}
where the functions $A(r),~B(r)$, and $C(r)$ are functions of the radial coordinate $r.$ Further, as $r\rightarrow\infty,~A(r),~B(r)\rightarrow 1.$ The matter tensors are described by the following equation:
\begin{equation}
T_{\alpha\beta}=(\rho+p_{t})u_{\alpha}u_{\beta}+p_{t}g_{\alpha\beta}+(p_{r}-p_{t})s_{\alpha}s_{\beta}\label{mat}
\end{equation}
where $u^{\alpha}=A^{-1}\delta^{\alpha}_{0},~s^{\alpha}=B^{-1}\delta^{\alpha}_{1}.$ 
Then the Einstein field equations $G_{\alpha \beta}=\kappa T_{\alpha \beta},$ for the above line element (\ref{metric1}) with the given matter tensors are obtained as:
\begin{align}
    \kappa \rho(r) &= \frac{1}{B^{2}} \left( \frac{B^{2}}{C^{2}}+\frac{2 B' C'}{B C}-\frac{C'^{2}}{C^{2}}-\frac{2C''}{C}\right) \\
    \kappa p_{r}(r) &= \frac{1}{B^{2}} \left(\frac{2 A' C'}{A C}+\frac{C'^{2}}{C^{2}}\right)-\frac{1}{C^{2}} \\
    \kappa p_{t}(r) &= \frac{1}{B^{2}} \left[ \left( \frac{A'}{A}-\frac{B'}{B}\right) \frac{C'}{C}-\frac{A' B'}{A B}+\frac{A''}{A}+\frac{C''}{C} \right],\label{fe}
\end{align}
where $'$ denotes the derivative with respect to the radial coordinate $r$ and $\kappa=8\pi G$ is the gravitational coupling term. 

In order to obtain an expression for complexity factor, following \cite{lh1} we redefine the matter tensors as follows:
\begin{equation}
T_{\alpha\beta}=\rho u_{\alpha}u_{\beta}+Ph_{\alpha\beta}+\Pi_{\alpha\beta}\label{mat1}
\end{equation}
where $P=\frac{p_{r}+2p_{t}}{3},~h_{\alpha\beta}=g_{\alpha\beta}+u_{\alpha}u_{\beta},~\Pi_{\alpha\beta}=\Pi(s_{\alpha}s_{\beta}-\frac{1}{3}h_{\alpha\beta})$ and $\Pi=p_{r}-p_{t}.$ We next consider the Misner-Sharp mass function $m(r)$, which includes the contribution due to the gravitational potential and kinetic energy, corresponding to a spherical shell of thickness $r-r_{0}$ in the above metric as \cite{ms}:
\begin{equation}
m(r)=m(r_{0})+\frac{\kappa}{2}\int_{r_{0}}^{r}\rho(\hat{r})C^{2}(\hat{r})C'(\hat{r})d\hat{r}\label{ms1}
\end{equation}
which on integration gives:
\begin{equation}
m(r)=\frac{\kappa}{6}C^{3}(r)\rho(r)-\frac{\kappa}{6}\int_{r_{0}}^{r}\rho'(\hat{r})C^{3}(\hat{r})d\hat{r}+M(r_{0})\label{ms2}
\end{equation}
where $M(r_{0})=m(r_{0})-\frac{\kappa}{6}C^{3}(r_{0})\rho(r_{0}),$ is the correction term due to the spherical shell. For any spherical distribution of matter, where $r_{0}=0$ we identically consider $M(r_{0})=0.$ This mass function will be useful in obtaining a suitable scalar expression for the complexity factor.

The Riemann curvature tensor can be expressed using the Weyl tensor $C_{\alpha\beta\mu\nu}$, the Ricci tensor $R_{\alpha\beta}$  and the Ricci scalar $R$ as follows:
\begin{equation}
R^{\nu}_{\alpha\beta\mu}=C^{\nu}_{\alpha\beta\mu}+\frac{1}{2}R^{\nu}_{\beta}g_{\alpha\mu}-\frac{1}{2}R_{\alpha\beta}\delta^{\nu}_{\mu}+\frac{1}{2}R_{\alpha\mu}\delta^{\mu}_{\beta}-\frac{1}{2}R^{\nu}_{\mu}g_{\alpha\beta}-\frac{1}{6}R(\delta^{\nu}_{\beta}g_{\alpha\mu}-g_{\alpha\beta}\delta^{\nu}_{\mu})\label{R1}
\end{equation}
For a spherically symmetric geometry without rotation the traceless Weyl curvature tensor reduces to the traceless, completely symmetric electric tensor given by: 
\begin{equation}
E_{\alpha\beta}=C_{\alpha\gamma\beta\delta}u^{\gamma}u^{\delta}\label{e1}
\end{equation}
The electric tensor can also be defined as $E_{\alpha\beta}=E(s_{\alpha} s_{\beta}-\frac{1}{3}h_{\alpha\beta}),$ where $E$ is defined in terms of the metric components as:
\begin{equation}
E=\frac{1}{2B^{2}}\left[\frac{A''}{A}-\frac{C''}{C}+\left(\frac{B'}{B}+\frac{C'}{C}\right)\left(\frac{C'}{C}-\frac{A'}{A}\right)\right]-\frac{1}{2C^{2}}\label{e2}
\end{equation} 
Using the field equations and (\ref{e2}) in an alternative definition of the mass function given by: $ m(r)=\frac{C^{3}}{2} R^{23}_{23}=\frac{C}{2}\left[1-\left(\frac{C'}{B}\right)^{2}\right]$ we get:
\begin{equation}
    m(r)=\frac{\kappa}{6}C^{3}(\rho-\Pi)-\frac{C^{3}E}{3}\label{ms3}
\end{equation}

We next consider the orthogonal split of the Riemann curvature tensor (first considered by Bel \cite{bel}) based on the Hodge duality, as follows: 
\begin{align}
Y_{\alpha\beta}&=R_{\alpha\gamma\beta\delta}u^{\gamma}u^{\delta}\label{yos}\\
Z_{\alpha\beta}&=*R_{\alpha\gamma\beta\delta}u^{\gamma}u^{\delta}=\frac{1}{2}\eta_{\alpha\gamma\epsilon\mu}R^{\epsilon\mu}_{\beta\delta}u^{\gamma}u^{\delta}\label{zos}\\
X_{\alpha\beta}&=*R^{*}_{\alpha\gamma\beta\delta}u^{\gamma}u^{\delta}=\frac{1}{2}\eta^{\epsilon\mu}_{\alpha\gamma}R^{*}_{\epsilon\mu\beta\delta}u^{\gamma}u^{\delta}\label{xos}
\end{align}
where `$*$' is the Hodge star operator and $R^{*}_{\alpha\beta\gamma\delta}=\frac{1}{2}\eta_{\epsilon\mu\gamma\delta}R^{\epsilon\mu}_{\alpha\beta}$ is the Hodge dual of the Riemann curvature tensor and $\eta_{\epsilon\mu\gamma\delta}$ the Levi-Civita tensor. Replacing the Ricci curvature and Ricci scalar from the field equation in (\ref{R1}) we get 
\begin{equation}
R^{\alpha\gamma}_{\beta\mu}=C^{\alpha\gamma}_{\beta\mu}+2\kappa T^{[\alpha}_{[\beta}\delta^{\gamma]}_{\mu]}+\kappa T\left(\frac{1}{3}\delta^{\alpha}_{[\beta}\delta^{\gamma}_{\mu]}-\delta^{[\alpha}_{[\beta}\delta^{\gamma]}_{\mu]}\right)\label{R2}
\end{equation}
From the above equation using (\ref{mat1}), $R^{\alpha\gamma}_{\beta\mu}$ can be split into into three distinct parts as:
\begin{equation}
R^{\alpha\gamma}_{\beta\mu}=R^{\alpha\gamma}_{I\beta\mu}+R^{\alpha\gamma}_{II\beta\mu}+R^{\alpha\gamma}_{III\beta\mu}
\end{equation}
From these three characteristic parts after some elaborate calculations ((follow \cite{lh1,lh2,lh3} for details)) we can get the following expressions for (\ref{yos})-(\ref{xos})
\begin{align}
Y_{\alpha\beta}&=\frac{\kappa}{6}(\rho+3P)h_{\alpha\beta}-\frac{\kappa}{2}\Pi_{\alpha\beta}+E_{\alpha\beta}\\
Z_{\alpha\beta}&=0\\
X_{\alpha\beta}&=\frac{\kappa}{2}\rho h_{\alpha\beta}-\frac{\kappa}{2}\Pi_{\alpha\beta}-E_{\alpha\beta}\label{os1}
\end{align}
From these tensors, four scalars can be defined of which the complexity factor $Y_{TF}$ is  considered in the current study and is obtained as: 
\begin{equation}
    Y_{TF}=-\frac{\kappa}{2} \Pi+E\label{ytf1}
\end{equation}
Substituting $E$  from (\ref{ms3}) and $m(r)$ from (\ref{ms2}) we obtain:
\begin{equation}
Y_{TF}=-\kappa\Pi+\frac{\kappa}{2C^{3}}\int^{r}_{r_{0}} \rho' C^{3}dr-\frac{3M(r_{0})}{C^{3}}\label{ytf2}
\end{equation}
with $M(r_{0})=0$ for $r_{0}=0.$ As is evident from the expression of $Y_{TF},$ it presents the pressure anisotropy, which is local and the mass inhomogeneity which is associated with the global properties of the object. For an isotropic and homogeneous system $Y_{TF}$ is found as zero, and such systems are classified as simple systems. The analogous representation of the complexity factor in terms of the metric tensors is:
\begin{equation}
Y_{TF}=\frac{1}{B^{2}}\left(\frac{A''}{A}-\frac{A'B'}{AB}-\frac{A'C'}{AC}\right)\label{ytf3}
\end{equation}
The expression (\ref{ytf2}) for the complexity factor is a slight modification of the original complexity factor expression suggested in \cite{lh1,lh2}. The modification is a result of the corresponding revision of the Misner-Sharp mass function corresponding to a spherical shell $(r-r_{0}).$ This facilitated the application of complexity factor for a wider class of space times like black holes, naked singularity or wormholes. In this article our aim is to discuss the complexity factor and concurrent occurrence of the photon spheres, hence it is imperative that we consider complexity of massive compact objects like the ones mentioned above which are known to be characterized by photon spheres and hence may be shadows.

\section{Photon Sphere and Shadows}

For static spherically symmetric objects described by (\ref{metric1}) the qualitative behaviour of the null geodesic at the equatorial plane $(\theta=\frac{\pi}{2})$ provide information on the effective potential. This effective potential is used to determine the radius at which photon sphere might exist for the gravitating object.

The Lagrangian of a particle confined to move along the equatorial plane of the considered metric is given by:
\begin{equation}
\mathfrak{L}=\frac{1}{2}\left[-A^{2}(r)\dot{t}^{2}+B^{2}(r)\dot{r}^{2}+C^{2}(r)\dot{\Phi}^{2}\right]\label{l}
\end{equation}
where {\it dot} represents differentiation wrt the affine parameter. From the above Lagrangian we obtain the constraints of motion as:
\begin{align}
E=A^{2}\dot{t}\quad\quad L=C^{2}\dot{\Phi}\label{cm}
\end{align}
Here $L$ is the angular momentum and $E$ is energy of the particle. Using (\ref{cm}) in (\ref{l}) we get the null geodesic $g_{\mu\nu}\dot{x}^{\mu}\dot{x}^{\nu}=0$ as:
\begin{equation}
\dot{r}^{2}=\frac{L^{2}}{A^{2}B^{2}}\left[\frac{1}{\mu^{2}}-V_{eff}\right]\label{em}
\end{equation}
where $\mu=\frac{E}{L}$ is the {\it impact parameter} and $V_{eff}=\frac{A^{2}}{C^{2}}$ is the {\it effective potential}. The critical radius $r_{c}$ of the null geodesic corresponds to $\dot{r}=0$ where we get $\mu(r_{c})=\frac{1}{\sqrt{V_{eff}(r_{c})}}=\frac{C(r_{c})}{A(r_{c})}.$ At the critical radius we get circular orbits for null geodesics if $\ddot{r}=0.$ These are called `light rings'. This translates to $V'_{eff}=0.$ Unstable circular orbits of the null geodesics (or light rings) corresponds to orbits of maximal potential or minimal period, which are called the {\it photon sphere}. The photon sphere also corresponds to the minimum value of the impact parameter $\mu.$ Thus photon spheres are obtained provided the $V_{eff}$ satisfies $V'_{eff}=0,~V''_{eff}<0$ at the critical radius $r_{c}=r_{ph}.$ If however $V'_{eff}=0,~V''_{eff}>0$ at $r_{ph}$ then we get a stable orbit null geodesic which is called the antiphoton sphere. Antiphoton spheres are mostly a mathematical curiosity and there is no physical evidence of its existence for asymptotically flat compact objects with regular event horizon. 

The angular deflection for a light ray travelling from infinity can be obtained as a function of the closest approach radius $r_{ca}$ as: 
\begin{equation*}
\alpha(r_{ca})=-\pi+2\int_{r_{ca}}^{\infty}\frac{B(r)}{C(r)\sqrt{\frac{V_{eff}(r_{ca})}{V_{eff}(r)}-1}}dr.
\end{equation*}
The above equation was obtained corresponding to a general metric in \cite{vir1} and for a Schwarzschild black hole space time in \cite{vir2}. Thus as $r_{ca}\simeq r_{ph}$ one can obtain the maximum deflection angle which forms the angular radius for shadow formation in case of massive gravity objects like black holes. The shadow radius is obtained from the minimum value of the impact parameter $\mu.$ 

\section{complexity and effective potential}

The effective potential $V_{eff}$ is the mathematical instrument to determine the photon spheres corresponding to spherically symmetric compact objects. Thus we want to connect the complexity factor $Y_{TF}$ with $V_{eff}$ so that we can obtain a quantitative estimate of how complexity impacts formation of photon spheres. In equation (\ref{ytf3}) we replace the function $A(r)$ using $C^{2}(r)V_{eff}=A^{2}(r)$ to obtain:
\begin{equation}
    Y_{TF}=\frac{1}{B^{2}}\left[\frac{1}{2}\left(\frac{V''_{eff}}{V_{eff}}\right)-\frac{1}{4}\left(\frac{V'_{eff}}{V_{eff}}\right)^{2}+\frac{C'}{2C}\left(\frac{V'_{eff}}{V_{eff}}\right)+\frac{C''}{C}-\left(\frac{C'}{C}\right)^{2}\right]-\frac{B'}{B^{3}}\left[\frac{1}{2}\left(\frac{V'_{eff}}{V_{eff}}\right)+\frac{C'}{C}\right]\label{yve}
\end{equation}
Given that $V'_{eff}(r_{ph})=0,$ at the photon sphere, the above expression reduces to:
\begin{equation}
    Y_{TF}(r_{ph})=\frac{1}{(B(r_{ph}))^{2}}\left[\frac{1}{2}\left(\frac{V''_{eff}(r_{ph})}{V_{eff}(r_{ph})}\right)+\frac{C''(r_{ph})}{C(r_{ph})}-\left(\frac{C'(r_{ph})}{C(r_{ph})}\right)^{2}\right]-\frac{B'(r_{ph})}{(B(r_{ph}))^{3}}\left(\frac{C'(r_{ph})}{C(r_{ph})}\right)\label{yph}
\end{equation} 
For further analysis we shall consider $C(r)=r$ which gives radial spherically symmetric objects.

\section{The Schwarzschild Black Hole Metric}

We consider a Schwarzschild black hole like metric by replacing $A(r)=\frac{1}{B(r)}$ as:
\begin{equation}
    ds^{2}=-B^{-2}(r)dt^{2}+B^{2}(r)dr^{2}+r^{2}(d\theta^{2}+sin^{2}\theta d\Phi^{2})\label{bh},
\end{equation}
where $B(r_{eh})\rightarrow\infty$ corresponding to the black hole event horizon at $r_{eh}<r_{ph}.$ Here the $V_{eff}(r)=\frac{1}{r^{2}B^{2}(r)}$ and $V'_{eff}=0$ gives $\frac{B'}{B}=-\frac{1}{r}.$ Using the above metric (\ref{yve}) reduces to
\begin{equation}
 Y_{TF}=\frac{r}{2}\left(rV''_{eff}+3V'_{eff}\right)
\end{equation}
so that at the location of the photon sphere we get
\begin{equation}
    Y_{TF}(r_{ph})=\frac{r_{ph}^{2}}{2}V''_{eff}(r_{ph})
\end{equation}
This clearly shows a direct correlation between the stability of the photon sphere and the complexity factor. We observe that
\begin{itemize}
\item $Y_{TF}>0\Leftrightarrow V''_{eff}>0:$ This gives $r_{ph}$ as the location of the antiphoton sphere.
\item $Y_{TF}<0\Leftrightarrow V''_{eff}<0:$ We get $r_{ph}$ the location of the photon sphere.
\item $Y_{TF}=0$ if $V''_{eff}=0:$ If $V''_{eff}=0$ at $r_{ph}$ then we cannot comment much about the space time. However, if $Y_{TF}=0$ for all values of $r$, then we essentially get $V_{eff}=\frac{\gamma}{r^{2}},~\gamma$ being a constant. This $V_{eff}$ can be obtained for the FRW metric, a generic zero complexity geometry, which is not characterized by any photon sphere.
\end{itemize}
This gives us a clear physical relevance of the complexity factor at $r=r_{ph}.$ If the complexity factor is negative at some radius $r_{ph}$ of a black hole then we can get a photon sphere at that location, whereas if it is positive then we can say that a photon sphere cannot be obtained at that location. If however the complexity factor reduces to zero at all values of $r$ then we can say that the region is far away from the influence of the black hole, such that the region is a part of the homogeneous and isotropic FRW space-time. 

\section{The static Morris-Thorne Traversable wormhole}

We now consider the static Morris-Thorne Traversable Wormhole (MTTW) space time, where $A^{2}(r)=e^{2\phi(r)}$ and $B^{2}(r)=\left(1-\frac{b(r)}{r}\right)^{-1}.$ Here the function $\phi(r)$ is the redshift function and the function $b(r)$ is called the wormhole shape function. This is because the shape of a wormhole is synonymous to its throat and the throat is identified at some radius $r=r_{0}$ where $b(r_{0})=r_{0}.$ It may be noted that the coordinate $r$ is not the actual radial distance from the throat, rather it is the radius of a circle around the throat $r_{0}.$ Thus $r\in (-\infty,r_{0})\cup(r_{0},\infty).$ The proper radial distance can be identified using the coordinate $l$ where $l$ is related to $r$ by the transform 
\begin{equation}
\frac{dl}{dr}=\pm\frac{1}{\sqrt{1-\frac{b(r)}{r}}},~l\in(-\infty,\infty)\label{tr}
\end{equation} and the throat being identified at $l=0.$ The wormhole shape satisfies the asymptotic flare out condition which is given by $b'(r_{0})<1.$ In order to determine the location of the photon sphere for this metric, we follow the procedure listed in section 3. We get $V_{eff}(r)=\frac{e^{2\phi(r)}}{r^{2}}$ in terms of $r.$ However $r$ being not the radial coordinate we need to transfer it to the actual radial coordinate $l$ using the transform (\ref{tr}). The location of the photon sphere is identified from \cite{wh6}
\begin{equation}
\frac{dV_{eff}(l)}{dl}=\left(\frac{dr}{dl}\right)\frac{dV_{eff}(r)}{dr}=\frac{2e^{2\phi(r)}}{r^{2}}\sqrt{1-\frac{b(r)}{r}}\left(\phi'(r)-\frac{1}{r}\right)=0.\label{dv}
\end{equation}
Thus there are two possible locations for appearance of light rings, one is at the throat $r=r_{0}$ while the other at $r_{ph}(>r_{0})$ which satisfies $\phi'(r_{ph})=\frac{1}{r_{ph}}.$ (In terms of the actual radial distance $l$, a $r_{ph}>r_{0}$ translates to two locations, $\pm l_{ph}$ obtained from the equation (\ref{tr}). Hence in terms of $l$ apart from the light ring at the throat such wormholes will have two more light rings, making the total light ring three). These light rings will be photon spheres provided $\frac{d^{2}V_{eff}(l)}{dl^{2}}<0.$ Now
\begin{equation}
\frac{d^{2}V_{eff}(l)}{dl^{2}}=\left(1-\frac{b(r)}{r}\right)\frac{d^{2}V_{eff}(r)}{dr^{2}}+\frac{1}{2}\left(\frac{b(r)}{r^{2}}-\frac{b'(r)}{r}\right)\frac{dV_{eff}(r)}{dr}\label{ddv}
\end{equation}
At the throat $r=r_{0},~b'(r_{0})<1$ which gives that the throat is a photon sphere provided $\phi'(r_{0})<\frac{1}{r_{0}}.$ Further $r_{ph}$ will be a photon sphere provided $\phi''(r_{ph})<-\frac{1}{r_{ph}^{2}}.$ Thus we see that a MTTW can have effectively more than one photon sphere. In case the wormhole have multiple photon spheres the shadow at the throat will be visually perceptible provided the potential function at the throat has a greater value than at the outer radius \cite{wh6}. 

To relate this to the complexity factor $Y_{TF}$ we rewrite the equation (\ref{yve}) by replacing $V'_{eff}$ and $V''_{eff}$ using $\frac{dV_{eff}}{dl}$ and $\frac{d^{2}V_{eff}(l)}{dl^{2}}.$ For the corresponding MTTW metric this gives:
\begin{equation}
Y_{TF}=\frac{\frac{d^{2}V_{eff}(l)}{dl^{2}}}{2V_{eff}}-\left(\frac{\frac{dV_{eff}}{dl}}{2V_{eff}}\right)^{2}+\sqrt{1-\frac{b(r)}{r}}\left(\frac{\frac{dV_{eff}}{dl}}{2rV_{eff}}\right)-\frac{b'(r)-\frac{3b(r)}{r}+2}{2r^{2}}\label{ywh}
\end{equation}
Thus at the location of the photon spheres we get:
\begin{equation}
Y_{TF}=\frac{\frac{d^{2}V_{eff}(l)}{dl^{2}}}{2V_{eff}}-\frac{b'(r)-\frac{3b(r)}{r}+2}{2r^{2}}
\end{equation}
We make the following observations from above:
\begin{itemize}
\item At the throat it may be noted that $\frac{b'(r)-\frac{3b(r)}{r}+2}{2r^{2}}$ is always negative. Thus if the throat is a photon sphere, that is if $\frac{\frac{d^{2}V_{eff}(l)}{dl^{2}}}{2V_{eff}}<0,$ then $Y_{TF}(r_{0})<0~(>0)$ depending on $\left|\frac{b'(r)-\frac{3b(r)}{r}+2}{2r^{2}}\right|<~(>)\left|\frac{\frac{d^{2}V_{eff}(l)}{dl^{2}}}{2V_{eff}}\right|.$ Conversely if $Y_{TF}<0$ then $\frac{\frac{d^{2}V_{eff}(l)}{dl^{2}}}{2V_{eff}}=Y_{TF}+ \frac{b'(r)-\frac{3b(r)}{r}+2}{2r^{2}}$ is always negative, which shows that if $Y_{TF}<0$ at the throat then a photon sphere will sufficiently exist at the throat but not necessarily. 

However, a $Y_{TF}>0$ can give rise to both, photon sphere or an antiphoton sphere depending on the mathematical behaviour of the functions at the throat. On the other hand if an antiphoton sphere exists at the throat then $\frac{\frac{d^{2}V_{eff}(l)}{dl^{2}}}{2V_{eff}}>0$ and then $Y_{TF}$ is necessarily positive at the throat. Thus existence of an antiphoton sphere sufficiently implies a positive $Y_{TF}$ but not necessarily.
\item Since $\frac{b'(r)-\frac{3b(r)}{r}+2}{2r^{2}}$ can be positive or negative at  $r=r_{ph}(>r_{0})$ no specific conclusion can be drawn connecting the existence or non-existence of the photon sphere with that of positive or negative complexity. However we can obtain a zero complexity at $r=r_{ph}$ if $\frac{\frac{d^{2}V_{eff}(l)}{dl^{2}}}{2V_{eff}}=\frac{b'(r)-\frac{3b(r)}{r}+2}{2r^{2}}$ at $r_{ph}.$ 
\item It may be noted that corresponding to a zero tidal force wormhole (constant red shift function) the $Y_{TF}$ vanishes for all $r$. For such wormholes the second light ring at $r_{ph}$ will not exist and a photon sphere will always exist at the wormhole throat. (Again we caution that existence of a photon sphere at the throat does not imply a zero complexity always or vice versa). 
\end{itemize}

Thus we observe that in case of a MTTW a negative complexity at the throat can be associated with a photon sphere at the throat, while an antiphoton sphere at the throat can mean a positive complexity.

\section{The Class of zero complexity Space-Time}

We consider all those generic class of space-times for which complexity vanishes. Zero complexity space times are considered as simple space times, which are equivalent to the homogeneous and isotropic FRW space. We have already seen that the zero-tidal force class of the MTTW is a zero-complexity space-time. In fact a quick look into  the equation (\ref{ytf3}) clearly shows that $Y_{TF}=0$ at all $r$ for any space time for which $A'(r)$ vanishes or $A$ is independent of $r.$ But there are other ways by which $Y_{TF}$ can vanish. Let us consider the equation (\ref{ytf3}) for $C(r)=r.$ Then $Y_{TF}$ can vanish for $B^{2}(r)=\alpha^{2}\left(\frac{A'(r)}{r}\right)^{2}$ with $\alpha$ being a constant. Thus any space-time of the form:
\begin{equation}
ds^{2}=-A^{2}(r)dt^{2}+\alpha^{2}\left(\frac{A'(r)}{r}\right)^{2}dr^{2}+r^{2}(d\theta^{2}+sin^{2}\theta d\Phi^{2})
\end{equation}
will essentially have vanishing complexity at all $r.$ Then is it possible that we find a physically viable space time of the above kind with photon sphere. Evidently the $V_{eff}$ for the above metric is given by $\frac{A^{2}(r)}{r^{2}}$ and $V'_{eff}=0$ gives $\frac{A'(r)}{A(r)}=\frac{1}{r}.$ If a photon sphere exists, then at the location of the photon sphere $r_{ph}$ we see that $g_{rr}=\frac{g_{tt}}{r^{4}}$ which clearly suggests this cannot be a black hole either. Thus a black hole or a wormhole with the above features does not exist. 

A reasonable solution can be found if we take $\frac{A'(r)}{r}=(1+R r^{2})^{n}$ a generalization of the Finch-Skea anasatz \cite{fs}, with the Finch Skea star obtained for $n=1.$ Here $R$ is a constant and $n$ is any real number. The above anasatz gives $A^{2}=\frac{(1+Rr^{2})^{2n+2}}{4R^{2}(n+1)^{2}}.$ Since $\alpha^{2}$ is arbitrary we make it 1 and redefining $\frac{dt}{2R(n+1)}=d\tau$ the above metric takes the form:
\begin{equation}
ds^{2}=-(1+Rr^{2})^{2n+2}d\tau^{2}+(1+R r^{2})^{2n}dr^{2}++r^{2}(d\theta^{2}+sin^{2}\theta d\Phi^{2})
\end{equation}
the solution being valid for $n\geq -\frac{1}{2}.$ For $n=\frac{1}{2},$ the anisotropic analogue of the Finch-Skea star can be obtained \cite{rb} while  for $n=-\frac{1}{2}$ a dark energy star is obtained. The matter tensors for $n=-\frac{1}{2}$ become homogeneous and isotropic with $p_{r}=p_{t}=-\rho=3R.$ Thus the resulting object is supported by dark energy energy fluid with an outward acting pressure for $R<0.$ For $R>0$ the energy density is negative and hence not viable. The physically relevant configuration obtained for $R<0$ can be identified as the interior of a Gravastar with $0\leq r<\frac{1}{\sqrt{-R}}$. The Finch Skea star has physical relevance in description of a neutron star while the exterior of the Gravastar has been called the regular black hole due to the absence of the event horizon with a thin shell at the boundary \cite{gs3,gs4,gs5,gs6,gs7,gs8,gs9,gs10,gs11,gs12,gs13,gs14}. 

Remarkably none of the configurations have photon spheres. So essentially any object that is simple either due to a constant red-shift or due to a metric configuration characterized by $B^{2}(r)=\alpha^{2}\left(\frac{A'(r)}{r}\right)^{2}$ cannot have external photon spheres. (One must be cautious with the zero tidal force MTTW, which has zero complexity with a photon sphere at the throat $r_{0}$. We are however talking about the existence of any photon sphere for $r>r_{0}$). This can be seen more strongly if we consider the energy conditions of the matter tensors at $r_{ph}.$ For zero complexity configurations obeying $B^{2}(r)=\alpha^{2}\left(\frac{A'(r)}{r}\right)^{2},$ the radial null energy conditions $\rho+p_{r}=V''_{eff}\frac{r_{ph}^{2}}{\alpha^{2} (A'(r_{ph}))^{4}}$ which clearly depends on the sign of $V''_{eff}.$ For photon sphere to exist at $r_{ph}$ the $V''_{eff}<0$ and hence should be supported by energy condition violating matter. Clearly photon spheres cannot exist for any configuration where $Y_{TF}=0.$

\section{General Space-Time harbouring black hole, naked singularity and wormhole}

\renewcommand{\arraystretch}{2.5} 
\begin{table}
\centering
\begin{tabularx}{\textwidth}{c|p{1cm}p{2cm}p{3cm}p{3cm}p{4.5cm}p{2cm}}
\hhline{=======}
&$~~q~~$&$~~~M~~~$& {\it Object}&{\it Photon Sphere} &$Y_{TF}$& {\it Shadow Radius}\\
\hhline{=======}
\multirow{5}{*}{$\lambda=0$}&\ -1&--&Minkowski space &--&0&--\\
\cline{2-7}
&\ 0&$M>0$&Schwarzschild BH&$r_{ph}=3M$&$-\frac{1}{9M^{2}}<0$&$~~~3\sqrt{3}M$\\
\cline{2-7}
&\ -3&$M<0$&Naked Singularity&--&$\frac{6M}{r^{3}}\left(1-\frac{2M}{r}\right)^{-4}<0$ for any $r$&$~~~2M$\\
\cline{2-7}
&$\geq 1$&$M>0$&Black Hole&$r_{ph}=(3+q)M$&$-\frac{(1+q)^{q}}{M^{2}(3+q)^{q+2}}<0$&$~~~M\sqrt{\frac{(3+q)^{3+q}}{(1+q)^{1+q}}}$\\
\cline{2-7}
&$<-3$&$M<0$&Naked Singularity&$r_{ph}=(3+q)M$&$-\frac{(1+q)^{q}}{M^{2}(3+q)^{q+2}}<0$&$~~~M\sqrt{\frac{(3+q)^{3+q}}{(1+q)^{1+q}}}$\\
\hhline{=======}
$0<\lambda<\frac{1}{2}$&~~0&$M>0$&Wormhole&$r_{ph}=\frac{3M}{1+\lambda}(>2M)$&$\frac{(1+\lambda)^{2}(3\lambda-1)}{9 M^{2}},~\geq 0$ for $\frac{1}{3}\leq \lambda<\frac{1}{2}$ and $<0$ for $0<\lambda<\frac{1}{3}$&$~~~\frac{3\sqrt{3}M}{\sqrt{(1+\lambda)^{3}}}$\\
\hhline{=======}
$\frac{1}{2}<\lambda<1$&~~0&$M>0$&Wormhole&$r_{ph}=2M$ (throat)&$\frac{1}{16\lambda M^{2}}$&$~~~\frac{2M}{\sqrt{\lambda}}$\\
\hhline{=======}
\end{tabularx}
\caption[One]{Corresponding to general Damour-Solodukhin space time we provide the various space times that can be obtained for different values of the parameters $q,M,\lambda$.}
\end{table}

\begin{figure}
\centering
\subfigure[]{\includegraphics[width=0.46\textwidth]{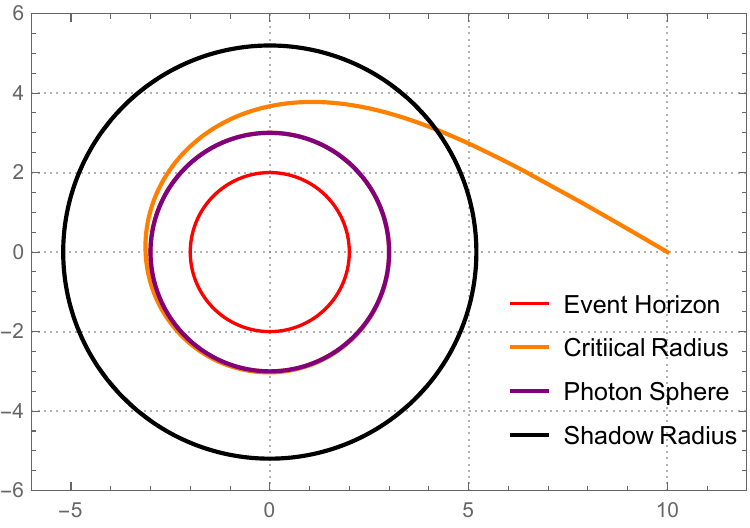}}
\subfigure[]{\includegraphics[width=0.46\textwidth]{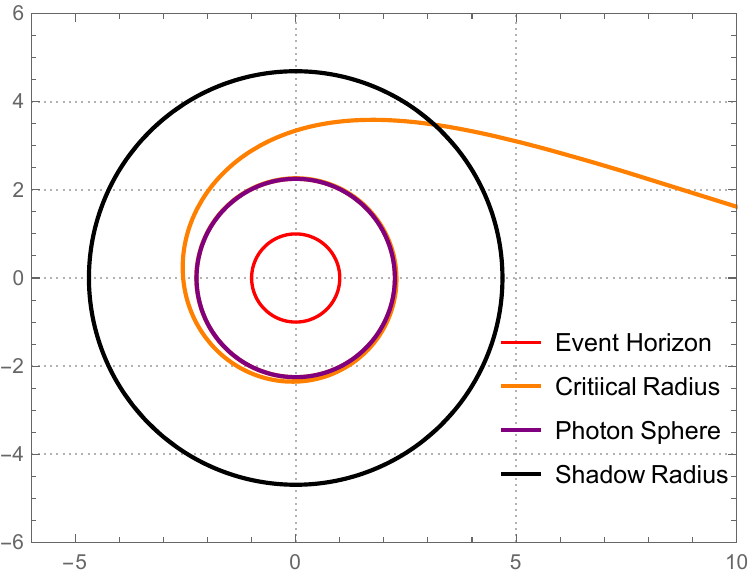}}
\subfigure[]{\includegraphics[width=0.46\textwidth]{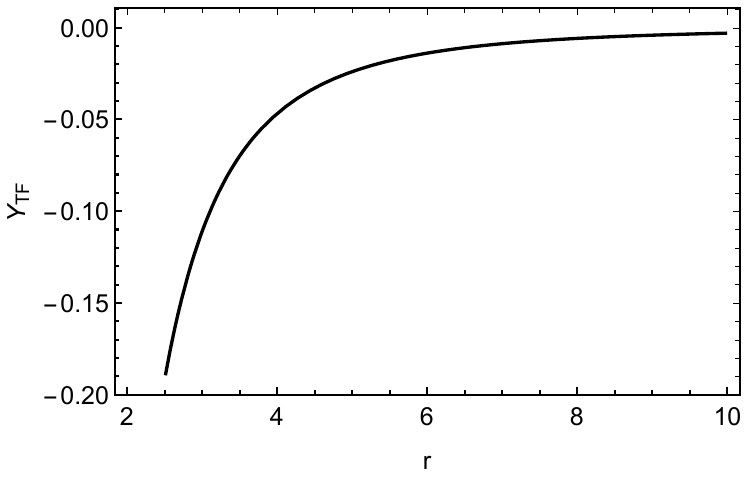}}
\subfigure[]{\includegraphics[width=0.46\textwidth]{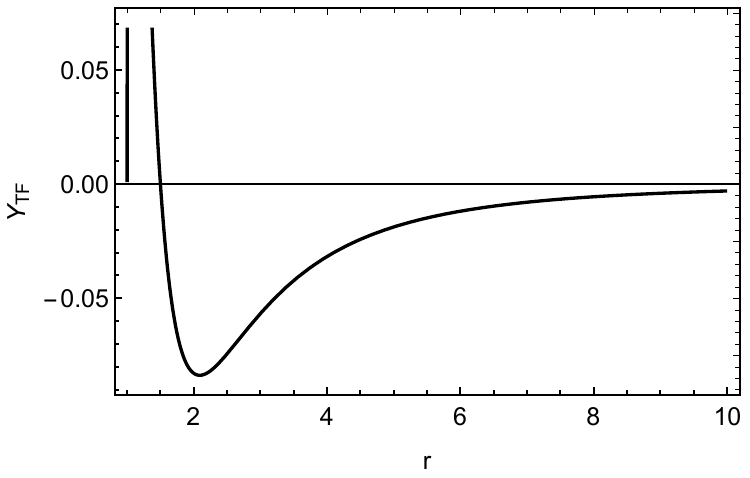}}
\caption{(a) This shows the light rings of the Schwarzschild black hole obtained for $M=1,~q=0.$ The photon sphere radius and event horizon radius is $r_{ph}=3,~r_{eh}=2$ respectively, while shadow radius is $3\sqrt{3}.$ (b) This shows the light rings of the black hole obtained for $M=0.5,~q=1.5.$ The corresponding photon sphere radius is $r_{ph}=2.25$, while event horizon occurs at $r_{eh}=1.$ The shadow radius is 4.691. (c) This shows the evolution of $Y_{TF}$ corresponding to the Schwarzschild black hole. Here $Y_{TF}<0$ for all values of $r$. (d) The evolution of $Y_{TF}$ corresponding to the black in sub-figure (b). Here $Y_{TF}<0$ at the photon sphere.}
\end{figure}

\begin{figure}
\centering
\subfigure[]{\includegraphics[width=0.46\textwidth]{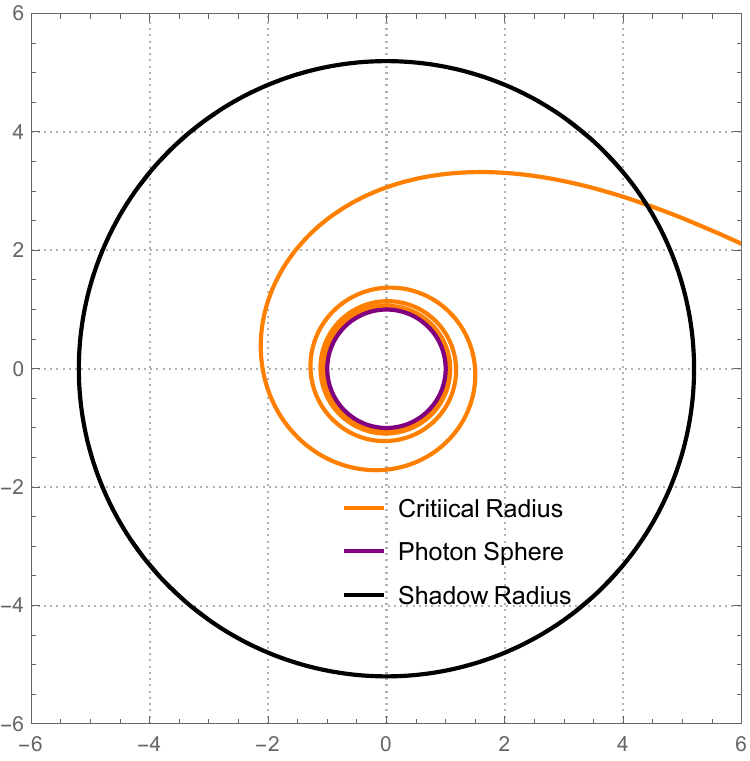}}
\subfigure[]{\includegraphics[width=0.46\textwidth]{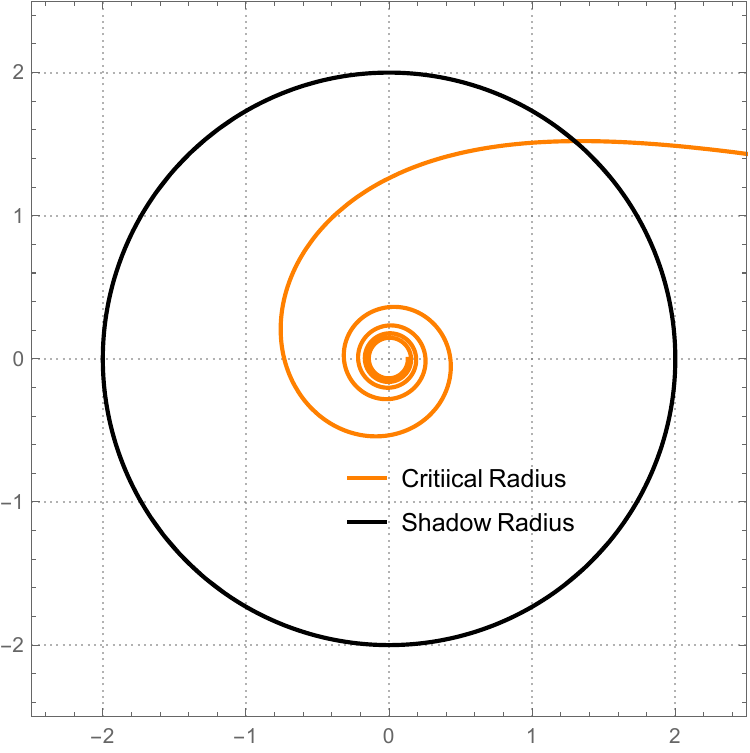}}
\subfigure[]{\includegraphics[width=0.46\textwidth]{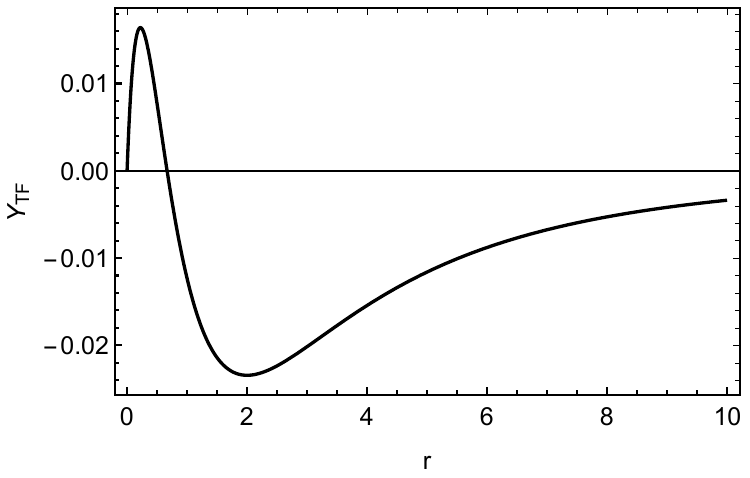}}
\subfigure[]{\includegraphics[width=0.46\textwidth]{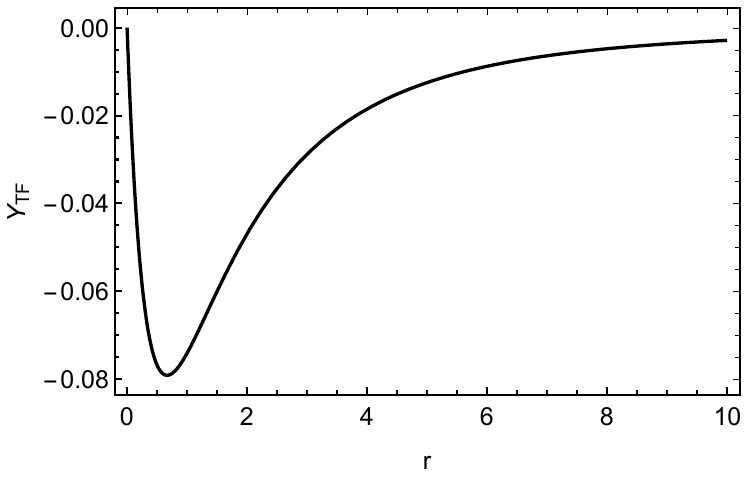}}

\caption{(a) This shows the light rings of the naked singularity obtained for $M<0,~q<-3.$ The figure has been drawn for $M=-1,~q=-4.$ Hence the photon sphere radius $r_{ph}=1$ and shadow radius is $3\sqrt{3}.$ (b) This shows the light rings of the naked singularity obtained for $M<0,~q=-3.$ The figure has been drawn for $M=-1.$ This however does not have a photon sphere and has a shadow radius 2. (c) The evolution of $Y_{TF}$ corresponding to the naked singularity obtained in sub-figure (a). One can see that $Y_{TF}<0$ at the photon sphere radius of 1. (d) The evolution of $Y_{TF}$ corresponding to the naked singularity obtained in sub-figure (b). Here $Y_{TF}<0$ for all values of $r$.}
\end{figure}

\begin{figure}
\centering
\subfigure[]{\includegraphics[width=0.46\textwidth]{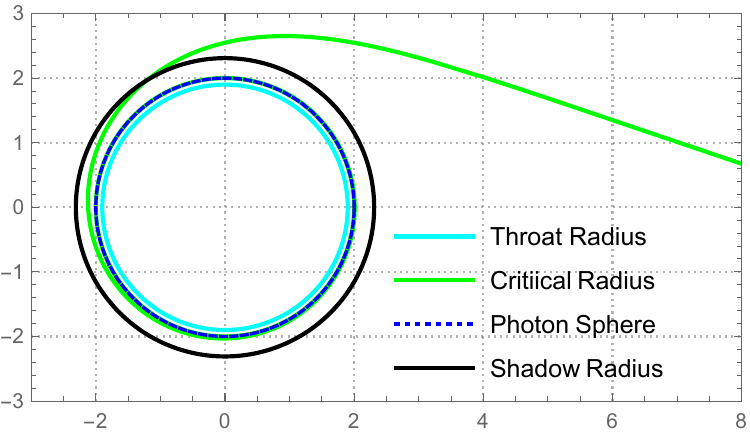}}
\subfigure[]{\includegraphics[width=0.46\textwidth]{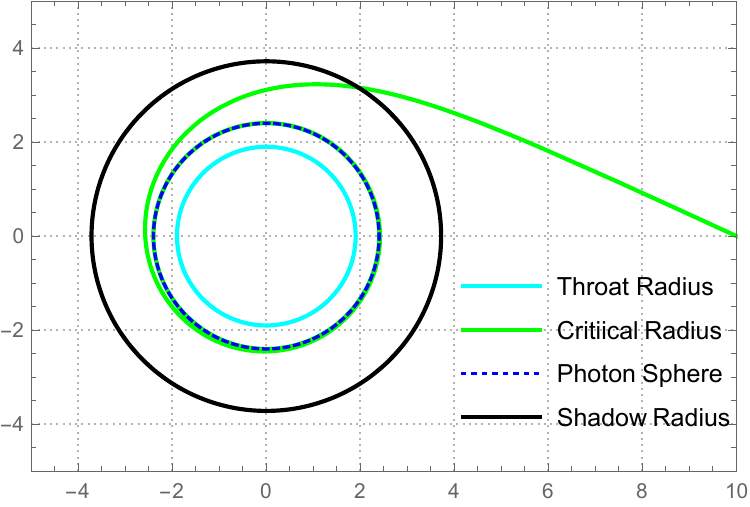}}
\subfigure[]{\includegraphics[width=0.46\textwidth]{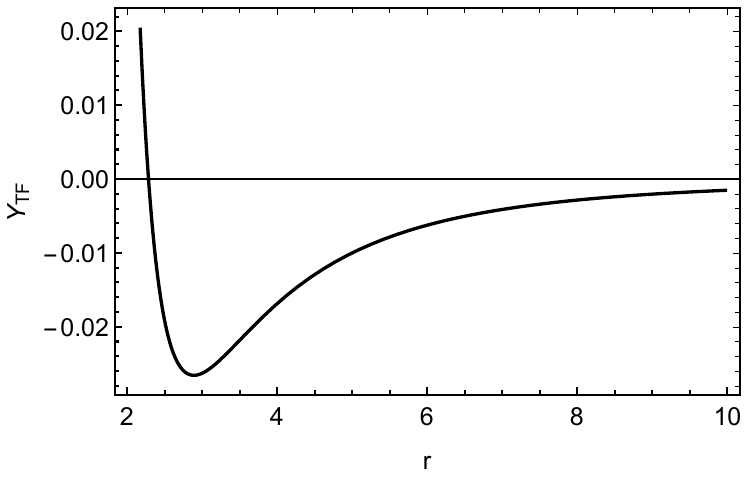}}
\subfigure[]{\includegraphics[width=0.46\textwidth]{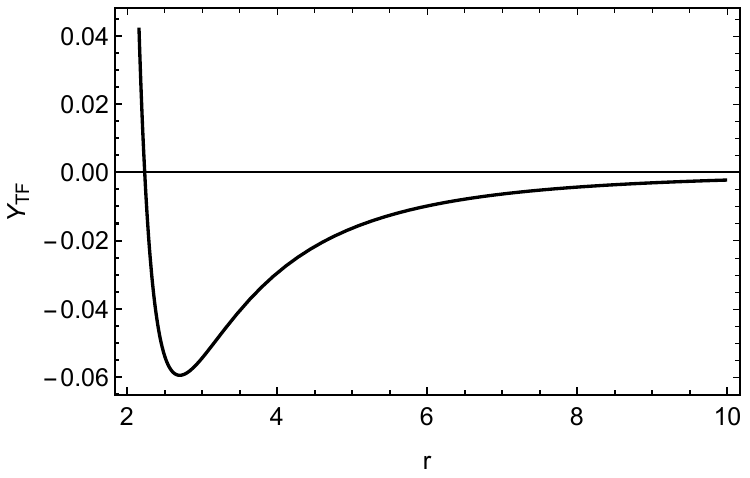}}

\caption{(a) This shows the light rings of the wormhole obtained for $M=1,~\lambda=\frac{3}{4}(>\frac{1}{2}).$ The throat radius is $r_{0}=2$ which is also the photon sphere radius $r_{ph}.$ The shadow radius is $\frac{4}{\sqrt{3}}.$ (b) This shows the light rings of the wormhole obtained for $M=1,~\lambda=\frac{1}{4}(<\frac{1}{2}).$ Here the throat is at $r_{0}=2,$ the photon sphere is obtained at $r_{ph}=\frac{12}{5},$ while shadow radius is at $\frac{24\sqrt{3}}{5\sqrt{5}}.$ (c) The evolution of $Y_{TF}$ corresponding to the naked singularity obtained in sub-figure (a). One can see that $Y_{TF}>0$ at the photon sphere radius of 2, which is also the throat location. (d) The evolution of $Y_{TF}$ corresponding to the wormhole obtained in sub-figure (b). Here $Y_{TF}>0$ at throat, which is an antiphoton sphere, while $Y_{TF}$ negative at photon sphere.}
\end{figure}

We consider the generalized Damour-Solodukhin like space time \cite{ds,dsw}, characterized by arbitrary parameters $q,~M(\neq 0)$ and $\lambda$ given by:
\begin{equation}
ds^{2}=-\left(1-\frac{2M}{r}+\lambda\right)^{1+q}dt^{2}+\left(1-\frac{2M}{r}\right)^{-(1+q)}dr^{2}+r^{2}(d\theta^{2}+sin^{2}\theta d\Phi^{2}),
\end{equation}
For $q=0$ and non-zero positive values of $\lambda$ we will get the Damour-Solodukhin wormhole. While for $\lambda=0$ and specific values of $q$ and $M$ this can be a naked singularity, a general black hole or the Schwarzschild black hole. We present this as an illustrative example for the results we proposed in the previous sections. 
The complexity of the above space time is given by:
\begin{equation}
Y_{TF}=\frac{M}{r^{3}}\left(1-\frac{2M}{r}\right)^{q}\left(1-\frac{2M}{r}+\lambda\right)^{-2}(1+q)\left[-\left(\frac{4M}{r}\right)^{2}(3+q)+\frac{M}{r}\left(12+7\lambda+q(2+\lambda)\right)-3(1+\lambda)\right]
\end{equation}
The effective potential $V_{eff}=\frac{\left(1-\frac{2M}{r}+\lambda\right)^{1+q}}{r^{2}}$ with $V'_{eff}=-\frac{2}{r}V_{eff}\left(\frac{1+\lambda-(3+q)\frac{M}{r}}{1-\frac{2M}{r}+\lambda}\right).$ In table 1 we give the different configurations for different values of the parameter.

The case for $q=-3,$ which gives a naked singularity without a photon sphere, but with a viable shadow radius has been discussed in details in \cite{ns4}.
We note that for the first wormhole configurations $0<\lambda<\frac{1}{2}$ three light rings exists, one at the throat and other two exterior to the throat (one has to be careful while considering the exterior to the throat, in a wormhole there are two exteriors one for $l>0$ and other for $l<0,$ this effectively means the existence of three light rings).  But the throat will form the antiphoton sphere and hence not mentioned in the table 1. Further we see that for the second wormhole only one light ring exists at the throat, which forms the photon sphere. (It may be noted that equations (\ref{dv}) and (\ref{ddv}) were used to find the photon spheres in the wormhole configurations with positive $\lambda.$ Curiously one may find that if $\lambda<0$ then again multiple photon spheres will exist, at the throat and also exterior to the throat at $\pm\frac{3M}{1+\lambda}$, however we neglect that because the corresponding metric will have negative red shift at the throat or a general geometry with hyperbolic symmetry.) 

Figures 1, 2 and 3 show the graphical representations of the black holes, naked singularities and wormholes discussed in table 1. Figure 1a shows the null geodesics of the Schwarzschild black hole with $M=1.$ We have plotted one geodesic corresponding to the critical radius $r_{c}.$ We shown the locations of the event horizon, photon sphere and shadow radius. Below that, in the same frame, figure 1c, shows the corresponding evolution of the complexity factor in a region exterior to the event horizon. As postulated, we see that at the photon sphere (where we get an unstable circular null geodesic with maximum $V_{eff}$) the complexity factor $Y_{TF}$ is negative. Figure 1b, shows the geodesics for the black hole obtained for $M>0$ and $q>1.$ The image has been drawn for $M=0.5$ and $q=1.5,$ and in figure 1d, we have plotted the corresponding $Y_{TF}$ in region outside the event horizon. As expected at the  photon sphere radius of $2.25$ we get negative $Y_{TF}.$ The graph of $Y_{TF}$ is also a profile of the evolution of $V''_{eff}$ which as can be seen attains a minimum at the photon sphere radius.

Figure 2a, 2b is corresponding to the naked singularity obtained for $q<-3$ and at $q=-3$ respectively. Here, it may be noted that, the metric representation of a naked singularity can be generalised from the black hole metric (\ref{bh}), with corresponding connection between complexity and photon sphere similar to that as obtained in case of a black hole. Although we did not explicitly postulated the connection, yet from the graphical representations in figures 2c and 2d, representing the corresponding complexity profile outside the singularity, one can see that $Y_{TF}$ is negative at the location of the photon sphere in 2c, while for the second case where there is no photon sphere, the complexity in general is negative.

Figure 3 is drawn for the two wormhole scenarios that could be obtained for $\lambda\neq 0.$ Figure 3a is corresponding to $\lambda>\frac{1}{2}.$ This wormhole has only one photon sphere at the throat, where $Y_{TF}$ is positive. As we had postulated, for a photon sphere at the throat, one could not necessarily comment on positivity of $Y_{TF},$ in this case it turns out to be positive at the throat radius $r_{0}=r_{ph}=2$ as can be observed from figure 3c. Figure 3b is obtained for $\lambda<\frac{1}{2}.$ In this case an antiphoton sphere exists at the throat $r_{0}$ while a photon sphere exists at an external radius outside the throat. As postulated, the existence of an antiphoton sphere at the throat will give a positive $Y_{TF},$ we observe from figure 3d that $Y_{TF}$ is positive at the throat location $r_{0}=2.$

\section{Discussion}

In this article we have successfully connected the complexity of a static spherically symmetric compact object with its light rings and stability. The complexity at the location of a light ring can in general tell us whether they will be stable or unstable.  Unstable light rings with maximum potential being identified with photon spheres and corresponding minimum impact gives us the shadow radius. It may be noted that in \cite{lh1,lh2} the final expression of complexity was given using the matter stress tensors. Here however we have exclusively used the geometric variables, that is the metric tensors to quantify our complexity, because these metric tensors are in general used to obtain the null geodesics and hence the light rings of the compact objects. The motivation of the article was to provide a physically relevant understanding of the mathematically relevant ``complexity factor" of a space-time. We have provided physical basis of positive, negative or zero complexity factor corresponding to certain physically relevant space-times. It may be noted that, the knowledge of complexity factor at some given radius is not the basis to draw conclusions on the existence of light rings or their stability. We have however, successfully provided the physical relevance of complexity of any space-time at some given radius. The underlying reason may be that, the complexity factor is intrinsically connected with the matter tensors not just through the Einstein's equations but also through the Misner-Sharp mass, which in turn gives the total energy of the system. As we know that the energy $E$ is useful in defining the light rings of a metric, it is therefore expected that complexity of the space-time will be related to the existence of light rings, its stability and hence shadow formation.

In \cite{lh1} zero complexity systems were found equivalent to the FRW space time, further zero complexity systems were earlier identified as stable for homologous, shear free and non dissipative systems \cite{lh4}. Here in this article we have identified such systems as ones where photon spheres or shadows cannot be obtained. This is probably because, as already identified, the fluid in a zero complexity system will evolve as a geodesic flow, which evidently will fail to exhibit geodesics similar to those necessary for obtaining photon spheres or shadows. Thus the otherwise purely theoretical explanation of complexity have now been related to physically relevant observables in the universe. Further we have provided a viable physically relevant geometry for vanishing complexity, which does not have a photon sphere. This raises relevant questions that whether it is possible to connect complexity with the black hole entropy or in general entropy of any compact object. It would be also interesting to investigate the complexity of rotating geometries and their impact on formation of innermost stable circular orbits (ISCO), photon spheres and entropy. We hope to investigate these as a future course of work.

\section{ACKNOWLEDGEMENTS:}
SN acknowledges UGC, Government of India, for financial assistance through junior research fellowship (NTA ref.no. 231610097492). SB acknowledges IUCAA, Pune, India, for hosting SB through their visiting associateship program, while working on the project.


\begin{thebibliography}{}
\bibitem{lh1}L. Herrera {\it Phys. Rev. D}, {\bf 97}, 044010 (2018).
\bibitem{lh2}L. Herrera, A.  Di Prisco and  J. Ospino,  {\it Phys. Rev. D}, {\bf 98}, 104059 (2018).
\bibitem{cf}L. Herrera, A. Di Prisco, J. Ospino, {\it Phys. Rev. D}, {\bf 99}, 044049 (2019) arXiv:1902.10133 [gr-qc];

L. Herrera, A. Di Prisco, J. Carot, {\it Phys. Rev. D}, {\bf 99}, 124028 (2019), arXiv:1906.08640 [gr-qc];

S. Khan, S. A. Mardan, M. A. Rehman, {\it Eur. Phys. J. C}, {\bf 79}, 1037 (2019);

M. Zubair, H. Azmat, {\it Int. J. Mod. Phys. D}, {\bf 29}, 2050014 (2020);

Z. Yousaf, M. Z. Bhatti, T. Naseer, I. Ahmad, {\it Phys. Dark Univ.}, {\bf 29}, 100581 (2020), arXiv:2004.14818 [physics.gen-ph];

M. Zubair, H. Azmat, {\it Phys. Dark Univ.}, {\bf 28}, 100531 (2020) arXiv:2012.14552 [gr-qc];

L. Herrera, A. Di Prisco, J. Ospino, {\it Eur. Phys. J. C}, {\bf 80}, 631 (2020), arXiv: 2007.12029 [gr-qc];

E. Contreras, E. Fuenmayor, {\it Phys. Rev. D}, {\bf 103}. 124065 (2021) arXiv:107.01140 [gr-qc];

Z. Yousaf, Kazuharu Bamba, M.Z. Bhatti, {\it Int. J. Mod. Phys. D}, {\bf 31}, 2250043 (2022);

C. Las Heras, P. Leon, {\it Gen. Rel. Grav.}, {\bf 54}, 138 (2022) arXiv:2203.16704 [gr-qc];

R. S. Bogadi, M. Govender, S. Moyo, {\it Eur. Phys. J. C}, {\bf 82},747 (2022);

Z. Yousaf, M. Z. Bhatti, M. M. M. Nasir, {\it Commun. Theor. Phys.}, {\bf 75}, 035401 (2023);

S. Bhattacharya, S. Nalui, {\it J. Math. Phys.}, {\bf 64}, 052501 (2023) arXiv:2304.08877 [gr-qc];

S.K. Maurya, et. al., {\it Phys. Dark Univ.}, {\bf 42}, 101284 (2023);

S. Das, M. Govender, R. S. Bogadi, {\it Eur. Phys. J. C}, {\bf 84}, 13 (2024);

L. Herrera, A. Di Prisco, {\it Phys. Rev. D}, {\bf 109}, 064071 (2024), arXiv:2404.04901 [gr-qc].
\bibitem{syn}J. L. Synge,  {\it MNRAS}, {\bf 131}, 463-466 (1966).
\bibitem{pt}D. N. Page and K. S. Thorne,  {\it ApJ}, {\bf 191}, 499-506 (1974).
\bibitem{lm}J. P. Luminet,  {\it Astron. Astrophys.}, {\bf 75}, 228-235 (1979).
\bibitem{bh7}R. Shaikh, P. Kocherlakota, R. Narayan, P. S. Joshi, {\it MNRAS}, {\bf 482}, 52-64 (2019), arXiv:1802.08060 [astro-ph.HE].
\bibitem{bh8}V. Perlick, O. Yu. Tsupko, {\it Phys. Reports}, {\bf 947}, 1-39 (2022).
\bibitem{bh6}M. Guerrero, et. al., {\it Phys. Rev. D}, {\bf 105}, 084057 (2022), arXiv:2202.03809 [gr-qc]
\bibitem{bh1}T. Hale, D. Kubiznak, J. Menšíková, {\it Phys. Rev. D}, {\bf 109}, 084061 (2024), arXiv:2401.16259 [gr-qc].
\bibitem{bh2}X-Q Li, H-Pg Yan, X-J Yue, S-W Zhou, Q. Xu, {\it JCAP}, {\bf 05}, 48(2024), arXiv:2401.18066 [gr-qc].
\bibitem{bh3}V. Vertogradov, Ali Övgün, {\it Phys. Lett. B}, {\bf 854}, 138758 (2024), arXiv:2404.18536 [gr-qc].
\bibitem{bh4}Y-Z Li, Zhenjiang), X-M Kuang, {\it Eur. Phys. J. C}, {\bf 84}, 271 (2024), arXiv:2401.07495 [gr-qc].
\bibitem{bh5}Y. Sekhmani, et. al., {\it Class. Quant. Grav.}, {\bf 41}, 81 (2024), arXiv:2407.20621 [gr-qc].
\bibitem{wh1}J. G. Cramer, et al., {\it Phys. Rev. D}, {\bf 51}, 3117 (1995).
\bibitem{wh2}T. Muller, {\it Phys. Rev. D}, {\bf 77}, 044043 (2008).
\bibitem{wh3}F. Abe, {\it ApJ}, {\bf 725}, 787 (2010).
\bibitem{wh4}N. Tsukamoto, T. Harada, K. Yajima, {\it Phys. Rev. D}, {\bf 86}, 104062 (2012).
\bibitem{wh5}P. G. Nedkova, V. K. Tinchev, S. S. Yazadjiev, {\it Phys. Rev. D}, {\bf 88}, 124019 (2013).
\bibitem{wh9}R. Shaikh, {\it Phys. Rev. D}, {\bf 98}, 024044 (2018) [1803.11422 [gr-qc]]. 
\bibitem{wh6}R. Shaikh, P. Banerjee, S. Paul, T. Sarkar, {\it Phys. Lett. B}, {\bf 789}, 270-275 (2019); {\it Erratum}, {\bf 791}, 422-423 (2019).
\bibitem{wh7}F. Rahaman, et. al. {\it Class. Quantum Grav.}, {\bf 38}, 215007  (2021) arXiv:2108.09930 [gr-qc]. 
\bibitem{wh8}R. Shaikh, P. Banerjee, S. Paul, T. Sarkar, {\it JCAP}, {\bf 07}, 028 (2019); {\it Erratum}, {\bf 12}, E01 (2023).
\bibitem{ns1}S. Paul, {\it Phys. Rev. D}, {\bf 102}, 064045 (2020).
\bibitem{ns3}S. Paul, R. Shaikh, P. Banerjee, T. Sarkar, {\it JCAP}, {\bf 03}, 055 (2020), arXiv:gr/qc 1911.05525.
\bibitem{ns4}A. B. Joshi, D. Dey, P. S. Joshi, P. Bambhaniya, {\it Phys. Rev. D}, {\bf 102}, 024022 (2020).
\bibitem{ns2}D. Dey, P. S. Joshi, R. Shaikh, {\it Phys. Rev. D}, {\bf 103}, 024015 (2021), arXiv:2009.07487 [gr-qc].

\bibitem{gs2}P. Mazur, E. Mottola, Gravitational vacuum condensate stars, {\it Proc. Natl. Acad. Sci.}, {\bf 101} 9545 (2004), arXiv:gr-qc/0407075].
\bibitem{gs16}N. Sakai, H. Saida and T. Tamaki, {\it Phys. Rev. D}, {\bf 90}, 104013 (2014), arXiv:gr-qc/1408.6929.
\bibitem{gs15}T. Kubo, N. Sakai, {\it Phys. Rev. D}, {\bf 93}, 084051 (2016) 

\bibitem{eht1}K. Akiyama, et al., (The Event Horizon Telescope Collaboration), {\it ApJL}, 875, L1-L6 (2019);

K. Akiyama, et al., (Event Horizon Telescope Collaboration), {\it ApJL} 930, L12-L17 (2022).
\bibitem{eht2}K. Akiyama, et al., {\it Astron. Astrophys.}, {\bf 681}, A79 (2024). 
\bibitem{sv}S. Vagnozzi, et al., {\it Class. Quantum Grav.}, {\bf 40}, 165007 (2023). 
\bibitem{go}G. J. Olmo1, J. L. Rosa, D. Rubiera-Garcia and D. Sáez-Chillón Gómez, {\it Class. Quantum Grav.}, {\bf 40}, 174002 (2023).
\bibitem{kp}Kunal. Pal, Kuntal Pal, R. Shaikh, T. Sarkar, {\it JCAP}, {\bf 11}, 60 (2023).
\bibitem{rs1}R. Shaikh, {\it MNRAS}, {\bf 523}, 375-384 (2023).
\bibitem{mg}Md. G. Mafuz, R. Diwan, S. Jana and S. Kar, {\it Eur. Phys. J. Plus}, {\bf 139} 219 (2024).
\bibitem{vd}V. Deliyski, G. Gyulchev, P. Nedkova, S. Yazadjiev, arXiv:2401.14092v1 [gr-qc] (2024).
\bibitem{nm} N. U. Molla, et. al., {\it Eur. Phys. J. C}, {\bf 84} 574 (2024);

N. U. Molla, H. Chaudhary, G. Mustafa, U. Debnath, S. K. Maurya, {\it Eur. Phys. J. C}, {\bf 84} 390 (2024).
\bibitem{nt}N. Tsukamoto, {\it Class. Quant. Grav.}, {\bf 40}, 228001 (2023),  arXiv: 2307.11303 [gr-qc].
\bibitem{ms}C. Misner and D. Sharp,  {\it Phys. Rev.}, {\bf 136}, B571 (1964).
\bibitem{bel} L. Bel, {\it C. R. Acad. Sci. Paris}, {\bf 247}, 1094-1096 (1958); {\it C. R. Acad. Sci. Paris}, {\bf 248}, 1297-1300 (1959); {\it Ann. Inst. H. Poincare}, {\bf 17}, 37 (1961).
\bibitem{lh3}L. Herrera, J. Ospino, A. Di Prisco, E. Fuenmayor and
O. Troconis, {\it Phys. Rev D}, {\bf 79}, 064025 (2009).
\bibitem{vir1}K. S. Virbhadra,  D. Narasimha, and  S. M. Chitre, Astron. Astrophys. 337(1998)1-8, arXiv: astro-ph/9801174.
\bibitem{vir2}K. S. Virbhadra and G. F. R. Ellis, Phys. Rev. D 62(2000)084003, arXiv: astro-ph/9904193.
\bibitem{fs}M. R. Finch and J. E. F. Skea, {\it Class. Quantum Grav.}, {\bf 6}, 467 (1989).
\bibitem{rb}R. Sharma, B. S. Ratanpal, {\it Int. J. Mod. Phys. D}, {\bf 22}1350074 (2013).
\bibitem{gs3}M. Visser and D. L. Wiltshire, {\it Class. Quant. Gravit.}, {\bf 21} 1135 (2004), arXiv:gr-qc/0310107].
\bibitem{gs4}C. Cattoen, T. Faber, M. Visser, {\it Class. Quant. Gravit.}, {\bf 22}, 4189 (2005), arXiv:gr-qc/0505137].
\bibitem{gs5}B. M. N. Carter, {\it Class. Quant. Gravit.}, {\bf 22}, 4551 (2005)
[gr-qc/0509087].
\bibitem{gs6}N. Bilic et al., {\it JCAP}, {\bf 0602}, 013, (2006), arXiv:astro-ph/0503427.
\bibitem{gs7}F. Lobo, {\it Class. Quant. Gravit.}, {\bf 23}, 1525 (2006). arXiv:gr-qc/0508115.
\bibitem{gs8}A. DeBenedictis et al., {\it Class. Quant. Gravit.}, {\bf 23}, 2303 (2006) [gr-qc/0511097].
\bibitem{gs9}F. Lobo et al., {\it Class. Quant. Gravit.}, {\bf 24}, 1069 (2007), arXiv:gr-qc/0611083.
\bibitem{gs10}D. Horvat et al., {\it Class. Quant. Gravit.}, {\bf 24}, 5637 (2007), arXiv:gr-qc/0707.1636.
\bibitem{gs11}B. M. H. Cecilia Chirenti and L. Rezzolla,{\it Class.
Quant. Gravit.}, {\bf 24}, 4191 (2007), arXiv:gr-qc/0706.1513.
\bibitem{gs12}P. Rocha et al.,{\it JCAP}, {\bf 11}, 010 (2008), arXiv:gr-qc/0809.4879.
\bibitem{gs13}K. K. Nandi et al., {\it Phys. Rev. D}, {\bf 79}, 024011 (2009), arXiv:gr-qc/0809.4143.
\bibitem{gs14}T. Harko, Z. Kova´cs and F. S. N. Lobo, {\it Class. Quant. Grav.}, {\bf 26}, 215006 (2009), arXiv:gr-qc/0905.1355.
\bibitem{ds}T. Damour and S. N. Solodukhin, {\it Phys. Rev. D}, {\bf 76},
024016, (2007), arXiv:0704.2667 [gr-qc].
\bibitem{dsw}K. K. Nandi, R. N. Izmailov, E. R. Zhdanov, A. Bhattacharya, {\it JCAP}, {\bf 07}, 027 (2018), arXiv:1805.04679 [gr-qc];

A. Ovgun, {\it Phys. Rev. D}, {\bf 98}, 044033 (2018), arXiv:1805.06296 [gr-qc];

R. K. Karimov, R. N. Izmailov, and K. K. Nandi, {\it Eur. Phys. J. C}, {\bf 79}, 952 (2019), arXiv:1901.05762 [gr-qc];

N. Tsukamoto, {\it Phys. Rev. D}, {\bf 101}, 104021 (2020), arXiv:2004.00822 [gr-qc].
\bibitem{lh4}L. Herrera, A. Di Prisco, J. Ospino, {\it Gen. Relativ. Gravity}, {\bf 42}, 1585–1599 (2010).


\end{thebibliography}
\end{document}